\documentclass[showpacs,preprintnumbers,amsmath,amssymb]{revtex4}
\usepackage {graphicx}
\usepackage{dcolumn}
\usepackage{bm}
\newcommand{\pmbhs}{P$\mu$BHs}
\newcommand{\pmbh}{P$\mu$BH}

\begin{document}
\preprint{}
\title{Stability of primordial black holes}

\author{Yoshiyuki Takahashi}

\affiliation{Department of Physics, The University of Alabama in Huntsville, Huntsville, AL 35899, USA.
\\
and\\
Computational Astrophysics Laboratory, RIKEN, Wako-shi, Saitama 351-0198, Japan.
}

\date{May 6, 2004}

\begin{abstract}
Vacuum polarization in a strong field, such as Hawking radiation from black holes, is constrained by the directional metric of strong field vectors. Relative localization of the quanta and the Schwarzschild radius is examined. 
Concept of thermal equilibrium of excited vacuum is critically analyzed and judged as dis-equilibrium, because virtual modes or particles are mutually independent. 
The result shows that primordial micro-size black holes (\pmbhs) cannot produce much Hawking radiation via fermions, bosons or thermal photons. Primordial black holes produced in the early epoch of the Big Bang should be very stable over Hubble time and consequently, may constitute a significant part of the cold dark matter in the structural formation of the universe. \pmbhs~should be gravitationally highly-condensed as missing mass around galaxies and galaxy clusters, and remain very difficult to observe. 
\end{abstract}


\maketitle

\section{Introduction}

Understanding the nature of dark matter as well as the missing mass around galaxies and galaxy clusters is one of the most serious and well-publicized problems in contemporary astronomy, cosmology, and particle physics. This paper provides a revision of the predominant quantum physical description of primordial black holes (PBHs), and proposes that PBHs should be considered in a renewed light as possible constituents of missing mass and of dark matter in general. In the past, massive neutrinos were thought to be the prospective candidates for constituting missing mass and hot dark matter. However, upon closer experimental examination of solar neutrinos, the mass of neutrinos became bound to a very small value on the order of 0.01-1 eV/c$^2$, rendering the massive neutrino hypothesis much less tenable. Many hypothetical alternatives have since been proposed with both baryons (dark stellar objects) and non-baryons. The former includes brown dwarfs, extended thin gases, stellar mass black holes and dead quasars. The latter includes axions, weakly interacting massive
 particles (WIMPs), super-symmetric particles, and a plethora of exotic hypothetical particles. 

The luminous baryonic mass density being observed for universe ($\rho_b$) is only 3\% of the critical mass density $\rho_c=3 h^2 / 8 \pi G$ required for the flat universe model, where H denotes the Hubble constant and G is the gravitational constant.  About $\sim$35\% of $\rho_c$ is accounted for by gravitational missing mass and is regarded as Cold Dark Matter (CDM) bound to galaxies and galaxy clusters. Dark energy is considered for up to the remaining 65\% of the critical mass. It has recently become popular due to observational inference of an accelerating universe model from the data analyses of receding supernovae and of small-scale fluctuations of cosmic microwave background radiation (CMBR) that also indicate the flatness of universe. Dark energy is a new name for the cosmological constant used in Lemaitre solutions of Einstein equations and in inflation models, and for the critical mass of the cosmology for the Einstein-de Sitter flat Universe. However, it has negative pressure ($P_{\Lambda} = -\rho_{\Lambda} < 0$) [or in a more original definition: negative mass density] and has a characteristic repulsive force. This is opposite to what is gravitationally required in order to account for the local missing mass in actual galaxies and Clusters. 
 
Primordial black holes (PBHs) have not been so seriously considered in these last few decades as candidates for either the galactic missing mass, or for dark matter in general. The reason for this vague status is due mostly to a belief that they should have evaporated very quickly after the Big Bang as a result of Hawking radiation [1], and therefore, that they can not be around us still. However, there is no observational evidence [2, 3] for the type of explosive gamma ray or radio [4] emission from the rapid decay of micro black holes which Hawking radiation mechanisms predict as being observable [5, 2]. Moreover, as shown below, a relevant distinction of quantum requirements for the metric of the Schwarzschild radius and the evaporation mechanism at a black hole event
 horizon reveals that Hawking radiation cannot significantly occur from PBHs, or from primordial micro-sized black holes
 (\pmbhs) in particular.

\section{Hawking thermal radiation}
The fundamental quantum nature of radiation from a vacuum in strong gravitational fields of BHs was first pointed out in a theoretical fashion by Hawking [1], with an intrinsic and unstated assumption that acceleration occurs uni-directionally. This radiation mechanism was then applied to primordial micro-size black holes (\pmbhs) without much concern about the size of the BH itself. However, the metric changes from a unidirectional field geometry to an omni-directional radial field when the Schwarzschild radius decreases to the size of a Compton wavelength ($\lambda$) of quanta; in this situation, the acceleration vector differs significantly from that of the unidirectional case.
Hawking gave the mass-loss formula from his thermal treatment [1] based on a unidirectional geometry without a categorical separation of the mass (M) of black holes,

\begin{equation}
\frac{dM}{dt}=10^{-26}M^{-2}\; [g s^{-1}]
\label{eq1}
\end{equation}

According to this, the lifetime of a black hole with mass ($M$), measured in units of grams ($M_g$),
 would consequently be:
\begin{equation}
\tau_{Hawking} = 3 \times 10^{-27} M_g^3 \; [s]
\label{eq2}
\end{equation}

This is the well-known basis for the concept that \pmbhs~with a mass less than $10^{14}$ g should have evaporated much earlier than the Hubble time ($\approx 3\times10^{16}$ s), during which the universe is believed to have evolved from the Big Bang to the present. Microscopic analysis of the metric of the \pmbhs~shows a different picture: \pmbhs~ do not evaporate as fast as what had been previously thought. The responsible mechanism we provide in this paper is a well-known, simple quantum requirement: it is, namely, the unidirectional length scale of one Compton wavelength, required for pair materialization from a virtual state in a vacuum. Hawking indeed cautioned in the end of his first paper [1] that his scheme was ignoring quantum fluctuations on the metric and that these might alter the picture. 

The original Hawking paper [1] considered the case of thermal black body radiation from a plasma of many particles at a thin circumference around the Schwarzschild radius ($R_S$). The area of this thin shell is compared to entropy ($S = (c^3 k/2 G \hbar) 4\pi R_S^2$ in Schwarzschild metric), and the temperature ($T$) is defined by a thermodynamical relationship, $T = E/S \propto Mc^2/4\pi R_S^2$, leading to temperature $kT =\hbar c^3/8\pi k G M$ and the luminosity $L = 4\pi R^2\sigma T^4\propto M^{-2}$. 
The temperature-equivalent Planckian spectrum was later verified by quantum field theories in curved spacetime [6, 7, 8]. However, there is a prerequisite in order for this thermal equilibrium scheme and Eqs. \ref{eq1}-\ref{eq2} to be valid: the Compton wavelength ($\lambda$) of radiated quanta has to be smaller than the hole radius ($R_S$) so that entropy can be defined for the black hole geometry [9]. Moreover, the established quantum physics principles and microscopic considerations of the cutoff sphere do not support the idea of \underline{thermal plasma} itself around the Schwarzschild radius of small BHs, because only one fermion of each species is allowed at any given time within a sphere of the Compton wavelength radius, which entirely engulfs a \pmbhs~when Schwarzschild radius ($R_S$) is smaller than $\lambda$ . If not radiated as a thermal black body, photons must originate from the electromagnetic annihilation of particle pairs. This process is negligibly small when the 
real-particle population of fermions or bosons is limited to one or much less (as described later).

We shall analyze the nature of virtual particles in particle picture. Before going into that description, we need to examine the approach from quantum field theory and the applicability of thermal concepts for vacuum states of wave modes or particles. This analysis includes the essential condition of equilibrium, which is founded on the basis of (1) {\it mutual exchange of energies} and (2) {\it the existence of a large number of particles (i.e. numerous enough to be statistically meaningful) over a significant time duration} that allows such exchanges. Vacuum states or virtual particles, though satisfying the second condition, have extremely too short a duration of real modes, and cannot satisfy the first condition. These bodies interact only with external gravity but not with other bodies, and should be regarded as so many free bodies in dis-equilibrium . 

\subsection{Support by quantum field theory}

The postulated mechanism of Hawking radiation was refined later by quantum field theory [6, 8], which used the curved spacetime of general relativity and Bogoliubov transformations [10] of the Fock representation [11] of the wave modes in a vacuum bounded by an event horizon. The boiling of vacuum states at the Schwarzschild radius has been considered by many to be a general consequence of the vacuum modes of a very strong field in curved spacetime; it occurs when a vacuum is geometrically cut off from the universe by the infinitely large plane of an event horizon so that a wave function does not propagate to one side of the universe. However, this geometry ignores the fact that the surface is actually spherical with finite radius, which does not cut off vacuum modes entirely from universe. The quantum field refinements [6, 7, 8] treat this actual 3D spatial geometry of the cutoff boundary for 3D (r,t) waves as if it were a 1-D wall, and as a uni-directionally accelerating configuration in space, irrespective of the size or the mass of a black hole. This goes against the basic notion that a curved spacetime configuration is supposed to be really used for all the spatial coordinate vectors of motion and gravitation. Consequently, these past schemes may yield a significant over-estimation in the mass-loss rate by a factor $\sim 1/M^2$. 

\subsection{"Thermal" concept of vacuum -- dis-equilibrium and independence}
The authentic (or standard) thermal interpretation of Planckian spectrum of quantum field theory is not free from imperfection. Despite a lot of sophisticated wave transformations being used in quantum field theoretical treatments, the whole scheme has more of a geometric nature in terms of fitting the wavelength ($\lambda$) modes of massless quanta (photons) to the size of the circumference ($\lambda_R=2 \pi R_S$) of the Schwarzschild sphere. 

Is it really heat that is generated by moving bodies in vacuum Other than virtual massive quanta, there are no real bodies or photons in negative energies. Temperature of vacuum $T_0$ is always zero in classical physics when there are no real bodies in motion. $T_0$ in quantum-field picture is different. 
It has a positive non-zero value ($T_0 = \sqrt{<E_e^2>}/k > 0$) due to the transient energy-square average of {\bf virtual electrons}. Because they are not real but in virtual states and the time is too brief to appear as real particles, they can not exchange energies with other virtual particles. In this sense, although they have non-zero (temperature-like) energy variance, they do not form any equilibrium state like that of an ordinary thermal body. They may have negative-energy Fermi motion, but this vanishes when they go back in vacuum states. Virtual particles are thus independent, and they are not really collectively and thermally connected with any neighbouring quanta. 

The field energy (conventionally interpreted as temperature) can indeed increase with decreasing $R_S$ until $R_S \sim \lambda_e/2 \pi = \lambdabar_e$. 
Gravitational fields are effective in terms of causing a uni-directional acceleration for a brief 
transient time, so long as $R_S \gg \lambdabar_e$. 
However, when $R_S \leq \lambdabar_e$, virtual electrons are no more effectively 
excitable by a gravitational acceleration of a strong field due to the 3-D random fluctuation 
of the motional vector of electrons relative to the field vector. Thus, vacuum quantum temperature ($T_e$) ceases to increase when $R_S$ reaches $\lambdabar_e$. Consequently, thermal emission of photons by 
black body saturates to a constant value ($\sigma T_{e0}^4$, with $k T_0 = m_e c^2$ = 511 keV), 
as discussed in 2.1 (as over-estimation by a factor of $\sim M^{-2}$). 
We will discuss more in detail in the later Section 7. 

The temperature ($T_m$) of different species of quanta with mass (m) at the black hole 
horizon is inversely proportional to m, $T_m \propto T_e (me/m)$ (see Appendix A). Virtual quanta with mass larger than that of electrons have a much lower temperature $T_m$ in the same gravitational field due to their heavier mass. The energy loss by their black-body radiation is much lower than their energy loss by electrons by a factor of $(me/m)^4$. 
In addition, even if we accept a sense of thermal picture, they have to be in dis-equilibrium 
(i.e., independent and remain at their own low temperature). For example, luminosity ($L_{Top}$) by the temperature of a top-quark (whose mass is $3.4 \times 10^5 m_e$) emits $1.3 \times 10^{22}$ times less energy than that ($L_e$) of electrons at $M_g = 10^{17}$ due to dis-equilibrium of the thermal 
reality of virtual fields. The largest of the emission rate by top-quarks is ($L_{Top}/L_e) = 3 \times 10^{-6}$ 
at $M_g = 7.1 \times 10^{11}$, for which the Compton wavelength of the top quark equals the Schwarzschild radius, and the top-quark
temperature saturates, too. Hence, temperature of higher-mass quanta cannot play a major role for the lifetime of \pmbhs~ in dis-equilibrium of vacuum temperatures.

\subsection{Gravitational potential energy}

Furthermore, as will be shown in Section 5, the initial assumption of total field energy being 
used by Hawking was not gravitational field energy: the total mass-energy $M c^2$ was used instead. 
This is approximately fine for large black holes, but is not accurate for small ones. 
This alone causes a large difference in temperature definition by a factor of M for 
\pmbhs. As a result, the mass-loss rate by black body luminosity ($\propto T^4$) 
in the Hawking interpolation scheme is extremely overestimated by a factor of $M^{-4}$
for small \pmbhs.

\section{Geometrical overview and limitation by Pauli blocking for electrons}
We examine the geometry first in light of quantum physics constraints. In a semi-classical Dirac-Schwinger  particle picture, a virtual electron-pair is in its negative energy state, and it can become on-shell only after gaining enough energy by (coherent) interactions over a volume defined by $\lambdabar_e^3$ and a duration $t =\lambdabar_e /c \approx 1.3 \times10^{-21}$ s.  We recall in the first place that there is an important quantum principle that strictly prohibits the rapid evaporation of sub-\pmbhs. Pauli blocking applies to Fermi-Dirac quanta within one Compton-wavelength 4-volume.  Any quantum fluctuation within one Compton-wavelength cannot produce any more than one electron ($10^{27}$ g) per $10^{-21}$ s.  A \pmbh~with a mass less than $10^{17}$ g has a Schwarzschild radius, $R_S =2GM/c^2$, less than  $\lambdabar_e \sim10^{11}$ cm. 
Only one electron can exist in 10$^{-21}$ s 
over its entire event horizon. This leads to the conclusion that there is in the \pmbh~horizon no plasma that can be 
regarded as a blackbody for radiation.  This strictly sets a \underline{very severe upperbound}
 on the mass-loss rate for \pmbhs~to 

\begin{equation}
-\frac{dM}{dt}\mid_{MAX} = m_e (m_e c^2 /\hbar) < 10^{-6}\; [g s^{-1}]
\label{eq3} 
\end{equation}

regardless of mass M (for $M < 10^{17}$ g).  The corresponding absolute minimum lifetime is, 

\begin{equation}
 \tau^{Pauli} >   10^6 M_g \;[s]
\label{eq4}
\end{equation}
which obviously gives a much longer lifetime than eq. (2) for any \pmbh.  Hawking's interpolation of eq. \ref{eq2} to small-mass black holes gave the lifetime for a Planck-mass ($10^{-5}$ g) black-hole on the order of $10^{-42}$ s; close to Planck time itself. The Pauli Principle, however, does not allow such an extremely short quantum evaporation time (at least for electrons), constraining it to be longer than $\sim$10 s.  This huge discrepancy between a major quantum physics principle and Hawking's original interpolation requires more critical scrutiny, particularly when considering the geometric size of the Schwarzschild radius. In fact, Bekenstein's articles [9] that preceded the Hawking radiation papers
 pre-constrained the validity of the entropy concept itself in terms of the surface area of an event
 horizon to $\lambdabar < R_S$ {\it so that entropy can be defined for quanta to fit into the hole } (Ref [7], p. 274). 
 The formulae used by Hawking (and almost all the later follow-up treatments by quantum field theories)
 generally used the energy content of the quantum modes, but their thermal treatment is not valid for \pmbhs; $\lambdabar > R_S$.

In a more analytical particle picture, a very strong acceleration and energy gain ($> 2 m_e c^2$) is required during
 a brief duration of less than the light-crossing time of sub-Compton wavelength $\epsilon \lambdabar$:

\begin{equation}
\Delta \tau_C = \epsilon \Delta \tau, \;(where \Delta \tau_C\equiv \lambdabar/c),
\label{eq5}
\end{equation}
in order for a virtual particle ($E = - mc^2$) to acquire the rest-mass energy ($+mc^2$). General relativity around
 the Schwarzschild radius has a lapse function of curved spacetime, 

\begin{equation}
\zeta \equiv (1-R_S/r)^{-1/2},
\label{eq6}
\end{equation}
for the Schwarzschild line element:  radial ($s=\zeta$dr) and temporal ($\tau=\zeta^{-1}$dt) coordinates in geodesics,
$ds^2=(1-\frac{R_S}{r})c^2dt^2 - \frac{dr^2}{(1-\frac{R_S}{r})} - r^2(d\theta^2+sin^2\theta d\phi^2)$.
 This provides the condition required for a critical acceleration 

\begin{equation}
a_C \geq c/\Delta \tau 
\label{eq7}
\end{equation}

where  $\Delta \tau$ denotes the proper time of a particle. The required acceleration time $\Delta \tau$ 
(equivalent to $\epsilon$ times the
 light-crossing time of the Compton wavelength in proper time $\Delta \tau_C$) is  $\zeta^{-1} \Delta t=\epsilon \lambdabar/c$ .  
Particle producing critical acceleration is possible in a thin shell surrounding the Schwarzschild
 radius with the condition that the particle is linearly localizable within this strong gravitational
 field and within the limited time. The length parameter ($\epsilon$) for the particle's location 
($r = R + \epsilon \lambdabar$ ) is
 restricted by the lapse function and the critical acceleration:  

\setcounter{equation}{6}
\begin{subequations}
\begin{equation}
a_C \geq c/\Delta t = \epsilon^{-1} \zeta^{-1} (c^2 / \lambdabar) = \epsilon^{-1} \zeta^{-1} a_0
\label{eq7prime}
\end{equation}
\end{subequations}

where
\begin{equation}
a_0 \equiv c^2/\lambdabar = 3.2 \times 10^{31} \;cm\; s^{-2} = 3.3 \times 10^{28} {\bf g} 
\label{eq8}
\end{equation}

with  {\bf g} = 980 cm $s^{-2}$ denoting the gravitational acceleration on the Earth's surface.

\section{Two domains of black hole mass for the quanta's Compton Wavelength}

Hawking radiation may occur just outside the event horizon but only within a spherical shell ($\Delta r < \epsilon \lambdabar$) and within 
a short period ($\Delta t \sim \epsilon \lambdabar/c$; $\epsilon < 1$).  If a virtual particle receives such a 
{\bf continuous unidirectional} strong
 gravitational force, it can be accelerated within $\Delta t$ to the energy of a real particle. This strong acceleration 
($a_S$) is characteristic of the lapse function of the radius ($r = R_S + \epsilon \lambdabar$) around the event horizon,

\begin{equation}
a_S(r)=\frac{G M /r^2}{\sqrt{1 - R_S/r}} = \frac{c^2 R_S}{2 (R_S + \epsilon \lambdabar)^{3/2} \sqrt{\epsilon \lambdabar}} 
\label{eq9}
\end{equation}

There is a clear distinction between two regimes of mass M (or R), separated by a Compton wave-length:
 
(i) Relatively-large black holes ($R_S \gg \lambdabar$), where gravity is unidirectional.

(ii) Micro-black holes ($R_S \ll \lambdabar$), where gravity is no longer unidirectional. 

\section{Gravitational field energy of black holes}

We will examine in the first place the energy content of a vacuum used in the Hawking scheme
for these two different mass regimes.
 The thermal bath, if it is ever generated by gravity,
 is the result of the gravitational field at $R_S$.  Its energy content in the spherical shell bounded by the radii
 ($R_S < r < R_S + \lambdabar$) is the most meaningful part in vacuum polarization, and it should not be the total mass
 energy ($M c^2$) of a black hole but its external gravitational field energy. It is evaluated by self-gravitating energy as:

\begin{equation}
E_g=\int^{\infty}_{R_S+\epsilon \lambdabar} M a_S ds = \int^{\infty}_{R_S+\epsilon \lambdabar} \frac{G M^2 dr}{r^2 (1-\frac{R_S}{r})}
=GM^2 ln (1+\frac{\epsilon \lambdabar}{R_S})=\frac{M c^2}{2} ln(1+\frac{2 G M}{c^2 \epsilon \lambdabar})
\label{eq10}
\end{equation}

This potential energy for $R_S \gg \lambda (\epsilon=1)$ deviates
(by a factor of ln(M)) from the 
 Bekenstein-Hawking
 scheme that used $E_g = Mc^2$.

However the potential energy for  case (ii) is $E_g = \frac{M c^2}{2} ln(1+\frac{2 G M}{c^2 \lambdabar})$ $\simeq \frac{G M^2}{\lambdabar}$.
Therefore, temperature $T= (G M^2/ \lambdabar)(4 \pi R_S^2 c^3 k/ 2 G \hbar) = c \hbar/ (8 \pi k G \lambdabar)$
becomes a constant, irrespective of the mass of the black holes.
This dramatically deviates from the Hawking temperature $T=c^3 \hbar/ 8 \pi k G M$ by a factor of M. 
Consequently, energy loss rate is $-dM/dt = S \sigma T^4 \propto M^{+2}$, significantly samller than the 
Hawking formula, Eq. \ref{eq1}, $dM/dt = S \sigma T^4 \propto M^{-2}$.
This alone already suggests that the lifetime of \pmbhs~($R_S < \lambdabar$) should be much, much longer than
the Hawking formula, even though the concept of temperature is unquestioned.

The above result is consistent with our basic results 
of the absolute bound constrained by Pauli blocking for fermions, Eqs. \ref{eq3} and \ref{eq4}. Further analysis of particle production from vacuum,
 as described in the following, confirms that this discrepancy for $R_S < \epsilon \lambdabar$
indeed represents a failure of the
 Hawking scheme to characterize accurately the lifetime of \pmbhs.

\section{Sub-Compton coordinate parameter}

The potential energy [$E_g = G M^2 ln (1 + R_S/\epsilon \lambdabar)$] 
depends on ($\epsilon$) as well as on the ratio $R_S/\lambdabar \equiv \xi$. 
Correspondingly, temperature (T) in the thermal picture of waves, if defined by $E_g/S$, 
depends weakly on $\xi/\epsilon$; namely, $T = [c \hbar ln (1 + \xi/\epsilon) / 8 \pi k]$.  
The energy loss rate in the thermal model for $R_S < \epsilon \lambdabar$ is $dM/dt = S \sigma T^4 = (4 \pi  M^2) \sigma
 (c^4 \hbar/ 8 \pi k)^4 ln^4 (1 +\xi/\epsilon)$, and is very different from the Hawking's thermal scheme, Eq. \ref{eq1}.
The valid accelerating region constrains the sub-Compton parameter ($\epsilon$) to be smaller 
than some value in a particle picture. We examine below the constraints on 
($\epsilon$) under the condition that sufficient acceleration of vacuum particles can occur.

\subsection{Regime (i): $R_S > \epsilon \lambdabar$ and a particle picture of large black holes}

The lapse function $\zeta \cong \sqrt{\frac{R_S}{\epsilon \lambdabar}}$ for a stellar-size black hole ($R_S \gg \lambdabar$) 
is large:  $\zeta^2 \approx 10^{-16} M_g$ for $\epsilon = 1$, and the 
acceleration of a particle is sub-critical for on-shell production unless  $\epsilon \ll 1$: 

\begin{equation}
a_S^{R_S > \lambdabar} = (\frac{c^2}{2}) \sqrt{\frac{1}{\epsilon_r \lambdabar R_S}} = 
(\frac{1}{2 \sqrt{\epsilon}}) \sqrt{\frac{\lambdabar}{R_S}} a_0 \approx (\frac{a_0}{2 \sqrt{\epsilon}}) \sqrt{\frac{10^{16}}{2 M_g}}
\label{eq11}
\end{equation}

This requirement for a sufficient acceleration $a_S > a_0$ restricts the coordinate coefficient $\epsilon < 10^{16} /8 M_g$, which is about unity for a black hole with mass $M_g$ around $10^{15}$ (g).  
For a stellar mass black hole ($M_g > 3\times 10^{33}$), this acceleration is so weak at $r = R_S + \lambdabar$ that the critical position coefficient $\epsilon$ must necessarily be limited to small values, $\epsilon < 10^{-18}$. 
The 3D position of a particle with such a small value of $\epsilon$   ($10^{-18}$ times
 sub-Compton for stellar-mass black holes) becomes extremely uncertain, but acceleration is nonetheless unidirectional and the integration $a_S \Delta t$ (=c) can provide the required energy for a virtual particle to achieve higher energy states. 
With the exception of the case where the black hole mass is around $10^{15}$ g, or $\epsilon \ll 1$,
particles remain sub-thershold for real and go back into virtual states,
contributing only to the variance of the vacuum energy ($T_0 = \sqrt{<E_e^2>}/k > 0$).
Equilibrium thermal states do not occur and massless photons cannot be emitted from an equilibriated thermal bath. 
Photons can come out, if at all, only as secondaries of bremsstarhlung 
[$dn/dp_{\gamma} \propto p_{\gamma}^{-1} \theta(p_e - p_{\gamma}) d p_{\gamma}$] directed only toward the horizon, 
while electrons are only accelerated before going back to virtual states 
The mass-loss rate is practically null, and therefore, it has to be less than Eq. \ref{eq1}.
[Quantum field theory does not use a particle picture, and mass-less particles (photons) 
can still be emitted at long wavelengths so long as the field is thermally excited by gravity of any 
strength. However, thermal equilibrium is unlikely to be an applicable concept for virtual states of vacuum particles.] (*)

\subsection{Regime (ii): $R_S > \epsilon \lambdabar$   and non-thermal description of micro black holes} 

In the case of a small black hole, (ii), $R_S \ll \lambdabar$, the lapse function tends toward unity like Newtonian gravity 
and it turns out by Eq. \ref{eq9} that the gravitational acceleration is sufficiently strong, but again, only for small
 values of  $\epsilon < 1/\sqrt{2}$.

\begin{equation}
a_S^{R_S < \lambdabar} = 2^{-1} \epsilon^{-2} (R_S /\lambdabar) a_0 \leq 2^{-1} \epsilon^{-2} a_0
\label{eq12}
\end{equation}

These metric conditions must be applied to derive the total effective volume where acceleration for vacuum materialization becomes sufficient.

The product of the vacuum polarization density ($\rho_4$), effective 3-volume ($V_3$) for acceleration, and the quantum tunneling probability (P) gives the total materialization rate ($\varpi$) of particle-pairs per unit time in the strong gravitational field of a black hole: 
\begin{equation}
\varpi = \rho_4 V_4 (c/\lambdabar) P \;[s^{-1}]
\label{eq13}
\end{equation}

The vacuum density ($\rho_4$) of virtual electrons is derived by Schwinger [12] by using a free Lagrangian, which gives $\rho_4^e \sim (3/4 \pi) (\lambdabar_e^4/c)^{-1} = 1.4 \times  10^{52} cm^{-3} s^{-1}$. To evaluate the total effective volume we now focus our attention to the case (ii).  The materialization is constrained by $a_S > a_C$  as

\begin{equation}
a_S=2^{-1} \epsilon_r^{-2} (R_S / \lambdabar) a_0 > \epsilon_r^{-1} a_0
\label{eq14}
\end{equation}

It follows that $\epsilon_r < R_S / (2 \lambdabar)< 0.5$ and $a_S > (2 \lambdabar/R_S)(c^2 /\lambdabar)=2 c^2/R_S = c^4/G M$. 
Correspondingly, the {\bf total} effective four-volume is reduced to

\begin{equation}
\rho_4 V_4 \equiv \rho_4 \Delta_{\epsilon} V_3 \Delta_{\epsilon} t = \epsilon^4 < (R_S/2 \lambdabar)^4 = 
(G M/c^2 \lambdabar)^4 < 1/16
\label{eq15}
\end{equation}

This signifies the fact, on one hand, that the linear acceleration does not achieve the required energy for
 materialization in outer sphere of $r > \epsilon_r \lambdabar + R_S$. 
On the other hand, the inner sphere $r <  \epsilon_r \lambdabar + R_S$ is not quite localizable for linear acceleration, and the materializable region in the case of (ii) $R_S < \lambdabar$ tends to null. We examine this point further in terms of tunneling probability.   

\section{Quantum limitation of unidirectional acceleration}

The probability of vacuum polarization for tunneling into on-shell state by a uni-directional field is 
e-fold as represented by the Gamow factor:

\begin{equation}
P \cong e^{-\frac{2}{\hbar} \int^{x_{+}}_{x_{-}} q(x) dx}
\label{eq16}
\end{equation}
where q(x) denotes the imaginary momentum,

\begin{equation}
q(x)=\sqrt{m^2 c^2 - (W - [m a_S x]^2)/c^2}
\label{eq17}
\end{equation}

The scalar integral $a_S x$ is not valid for omni-directionally and randomly fluctuating quanta 
within $x < \lambda$, and Eq.\ref{eq17} should be written in its proper vector form,

\setcounter{equation}{16}
\begin{subequations}
\begin{equation}
q(x) = \sqrt{ m^2 c^2 -(W- [ \int m a_s \bullet d x]^2)/c^2}
\label{eq17b}
\end{equation}
\end{subequations}

This clarifies the fact that the total gravitational work (scalar product) of isotropically fluctuating
 two vectors is null:

\begin{equation}
\int a_s \bullet d x = \int \int^{+2 \pi}_{-2 \pi} a_S dx cos \theta \frac{1}{\sqrt{2 \pi} \sigma_{\theta}} e^{-\frac{\theta^2}{2 \sigma_{\theta}^2}} d\theta = 0
\label{eq18}
\end{equation}

Provided that the linear approximation, $a_S(x) \bullet x = a_S x$, is allowed for the integrand of the
 tunneling penetration factor, we get an imaginary upperbound of tunneling probability for fermions ($P_f$),

\begin{equation}
P_f \cong e^{-\frac{2}{\hbar c} \int_{x_-}^{x_+}\sqrt{m^2 c^4 - (W - m a_S x)^2/c^2} dx} =
e^{-\frac{2}{\hbar c} \frac{m^2 c^4}{m a_S} \int^{+1}_{-1} \sqrt{1 - u^2} du} 
\label{eq19a}
\end{equation}

\setcounter{equation}{18}
\begin{subequations}
\begin{equation}
P_f \cong e^{-\frac{\pi m c^3}{a_S \hbar}} \ll 1
\label{eq19b}
\end{equation}
\end{subequations}

The total emission rate per second by tunneling electrons from the vacuum that surrounds a \pmbh~is 

\begin{equation}
\varpi_{f}^{R_S} \leq (\frac{G M}{c^2})^4 (\frac{m c^2}{\hbar}) \varpi_f \;[s^{-1}]
\label{eq20}
\end{equation}

where  $\varpi_f$ denotes the particle creation rate per unit volume and time derived by the effective Lagrangian 
 ($L^{\prime}$) of quantum field theory for a unidirectional gravitational acceleration ($a_S$) field,

\begin{equation}
\varpi_{f} = 2 Im(L^{\prime}) = \frac{1}{4 \pi^3} (\frac{m a_s \hbar}{m^2 c^3})^2 \frac{m c^2}{\hbar} (\frac{m c}{\hbar})^3
\Sigma^{\infty}_{n=1} \frac{1}{n^2} e^{-n \frac{\pi m c^3}{a_S \hbar}} 
= \frac{1}{4 \pi^3} (\frac{m^2 a_S^2}{c \hbar^2}) \Sigma^{\infty}_{n=1} \frac{1}{n^2} e^{-n \frac{\pi m c^3}{a_S \hbar}}
\label{eq21}
\end{equation}

\setcounter{equation}{20}
\begin{subequations}
\begin{equation}
\leq \frac{1}{4 \pi^3} (\frac{1}{c \hbar^2}) (\frac{m c^4}{G M})^2 \Sigma^{\infty}_{n=1} \frac{1}{n^2} e^{-n \frac{\pi G M m}{c \hbar}}
\label{eq21a}
\end{equation}
\end{subequations}

The steep exponential factor as a function of mass ($m$) of quanta in Eq. \ref{eq21} assures that high-mass
 quanta are produced negligibly less than the low-mass quanta. Each term of the sum of the exponential 
series (called the Spence Function) represents n-loop coupling and is very small, unless acceleration 
is higher than the critical value $a_C$.  Thus, the total particle emission rate per second is bounded by,	

\begin{equation}
\varpi^{R_S}_{f} \leq (\frac{G M}{c^2})^4 (\frac{1}{c}) (\frac{m c^2}{\hbar}) \frac{1}{4 \pi^3} (\frac{1}{c \hbar^2})
(\frac{m c^4}{G M})^2 \Sigma^{\infty}_{n=1} \frac{1}{n^2} e^{-n \frac{\pi G M m}{c \hbar}}
\label{eq22a}
\end{equation}

\setcounter{equation}{21}
\begin{subequations}
\begin{equation}
\varpi^{R_S}_{f} \leq \frac{1}{4 \pi^3} (G M)^2 (\frac{m^3}{\hbar^3})  \Sigma^{\infty}_{n=1} \frac{1}{n^2} e^{-n \frac{\pi R_S}{2 \lambdabar}} 
< \frac{c}{16 \pi^3 \lambdabar} (\frac{R_S}{\lambdabar})^2 < \frac{c}{16 \pi^3 \lambdabar}
\label{eq22b}
\end{equation}
\end{subequations}
which is consistent with the $\varpi$  obtained earlier by the volume analysis: Eqs. \ref{eq13} and \ref{eq15}.

The production probability of bosons ($P_b$) in a strong field resembles that of fermions [13], with 
a coefficient smaller by only a factor of two for the Spence function.  

\begin{equation}
P_b \cong 2^{-1} \times e^{-\frac{\pi m c^3}{a_S \hbar}}
\label{eq23}
\end{equation}

\begin{equation}
\varpi_b = \frac{1}{8 \pi^3} (\frac{m^2 a_S \hbar}{c}) \Sigma^{\infty}_{n=1} \frac{1}{n^2} e^{-n \frac{\pi m c^3}{a_S \hbar}}
\label{eq24}
\end{equation}

The most critical problem for a tunneling integral for $x < \lambdabar$  is that the coordinate (x) fluctuates randomly over the path integral, changing the direction of the gravitational vector $a_S(x,t)$ {\bf so that it is totally uncertain} for the effective one-directional path integral of $a_S \bullet dx = a_S (dx) cos \theta$,  due to the uncertainty of $\theta$ $(-2 \pi \leq \theta \leq 2 \pi)$ at any time and position during this integration. This makes the Gamow factor practically zero for a black hole whose RS is smaller than the particle's Compton wavelength $\lambdabar$:

\begin{equation}
P\ll e^{-\frac{\pi m c^3}{a_S \hbar}}
\label{eq25}
\end{equation}

Thus, the 3-D uncertainty for the 1-D unidirectional integral in Eqs. \ref{eq17} and \ref{eq18} entirely kills the Gamow factor and makes the probability effectively zero as shown in Eq. \ref{eq19b}. It is the same situation as that of a strong electric field  $U(r) = Z e^2/4 \pi \epsilon r$ inside a uranium nucleus, e.g., ($r < 1$ fm), where the effective coupling probability becomes  $> 137 \alpha = 1.0$.   Literally, the vacuum could have decayed in a high electric field $Z e^2/r$ by a singular multiple coupling probability, if the unidirectional tunneling were ever allowed and if the Uncertainty Principle that requires $r > \lambdabar$  were ignored.  The case of \pmbh~vacuum polarization by strong gravity has the same problem as this uranium electric case.  The 1-D integral in the Gamow factor does not yield any meaningfully small value for the exponent within $r < \lambdabar$, and hence no spontaneous vacuum decay can take place, because \pmbhs~require Eq. \ref{eq14}, namely,   $\epsilon_r < R_S (2 \lambdabar) < 0.5$. 

For large BHs with $r \gg \lambdabar$, 1-D directionality is sufficient, and Eqs. \ref{eq11}, \ref{eq22b} and \ref{eq24} can be used. We recall Eqs. \ref{eq22b} and \ref{eq24} as the upperbound for energy loss from a BH.  The probability of acquiring $a_S > a_C$ is, however, very small for vacuum electron pairs with any gravitational acceleration in curved spacetime (for the \pmbhs~case) or electrostatic acceleration in flat spacetime (as in the uranium case where $r \gg 1$ fm), with the metric more than one-Compton wavelength away from the Schwarzschild radius or from a nucleus center, respectively. Thus, we are led to conclude that fermions or bosons can barely be emitted from \pmbhs. Consequently, no thermal radiation can occur from such a site where far less than one particle at any time can exist in the quantum-materialization spacetime volume.

\section{Mass-loss rate and half lifetime}
The mass-loss rate of \pmbhs~($M < 10^{17}$ g) can be evaluated as an upperbound by using Eqs. \ref{eq11} and \ref{eq22a} with P =1, but it is indeed an extreme overestimation in the sense that we 
ignore the ineffectiveness (Eq. \ref{eq25}) of a path integral in Gamow factor P $\ll$1.  It leads to an artificially faster evaporation rate; yet, this rate already gives many orders of magnitude less than that interpolated  by Hawking's thermal scheme.  Our upperbound is thus

\begin{equation}
-\frac{dM}{dt} \ll   m_e (\varpi^{R_S}_f +\varpi^{R_S}_b) \;[g s^{-1}] =
 3 \times 10^{-47} M_g^2 \;[g s^{-1}]
\label{eq26}
\end{equation}

The half lifetime ($\tau_{1/2}$) of a \pmbh~($M \leq 10^{17}$ g) in this scheme is 

\begin{equation}
\tau_{1/2 (P\mu BH)} \gg 3 \times 10^{48} M_g^{-1}\; [s]
\label{eq27}
\end{equation}

This suggests $\tau_{1/2} > 10^{31}$ s at $M \sim 10^{17}$ g ($R_S = \lambdabar$), which is consistent with the absolute 
lower bound for fermions ($\tau_{1/2}> 10^{23}$ s).  Any \pmbhs~with mass less than $10^{17}$ g can live 
many orders of longer time than the Hubble time ($\sim 3 \times 10^{16}$ s). The smallest \pmbh~is that 
of the Planck mass, ($\sim 10^{-5}$ g), which can live $\gg 10^{47}$ yr, and can provide a basis for a finite size spacetime without decay.

Consideration of the smallest-mass quanta represents a significant departure from the Hawking scheme of Eq. \ref{eq2}. We used electrons for its known firmness of the lightest-mass electromagnetic quanta in this physical world. Neutrinos have been treated as massless until recently, but they are supposed nowadays to have small finite masses 0.01 -1 $eV/c^2$ as inferred from the observed deficits of fluxes of solar electron neutrinos and atmospheric muon neutrinos. With these finite masses ($m_{\nu}$), neutrinos cannot be emitted from black holes with a mass smaller than the neutrino critical mass $M_C^{\nu} = 10^{23}$  (1 $eV/m_{\nu} c^2$) g. Other light-mass particles such as axions ($m_A < 10^{-14}$ eV) are hypothetical. They would make the thermal equilibrium basis of Hawking formula partially invalid for $M_C^A [< 10^{37} g\; (\sim 10^{4} M_{\odot})]$, if they really exist. Although these are fundamental quanta, they are neutrals and do not directly couple to photons. These quanta cease to contribute to the thermal bath of a vacuum for $M < M_C^{\nu} < M_C^A$  long before electrons do so at $M < M_C^e = 10^{17}$ g. 
The role of quanta at higher masses is negligible so long as they remain only as virtual states in vacuum, as already stated in 2.2, due to the disequilibrium nature of virtual particles.
The largest mass of fundamental quanta presently-known for production is that of the top quark (174 $GeV/c^2$) [14], which gives its critical mass $M_C^{top} = 7.6 \times 10^{11}$ g.  
Energy loss for a BH with a mass less than $10^{17}$ g (and above the critical mass $M_C^{top} = 7.6  \times  10^{11}$) can still take place.  The probability of pair production from vacuum 
fluctuation for such heavy particles is, however, much smaller than for electrons for $M > M_C^{top}$ due to the exponential factor in tunneling probability. The thermal bath of a vacuum for photon emission is correspondingly highly limited.  Similarly, due to the more stringent quantum conditions that destroy the Gamow factor by 3D fluctuations for $R_S (M) < \lambdabar$ (Eqs. \ref{eq19a} and \ref{eq25}), any unknown quanta with a mass heavier than the top-quark mass, if any would be discovered in future, can still safely be ignored in the consideration of the lifetimes of \pmbhs. 

\section{Conclusions and discussions}
In summary, the radiation rate from \pmbhs~is evaluated. Several different considerations and analyses indicated that the Hawking radiation from \pmbhs~ was substantially overestimated by a factor of 1/$M^2$ in the Hawking's thermal scheme with or without quantum field theoretical approaches. Those aspects having been analyzed are (1) metric in the original Hawking scheme, (2) geometry in quantum field theoretical approaches, (3) gravitational energy for small black holes, (4) non-unidirectional tunneling probability, and (5) dis-equilibrium nature of virtual quanta.

We are now led to conclude that all the \pmbhs~should be regarded as very massive, cold, and dark (black) matter. They would not have been lost since their generation in the early big-bang era, were \pmbhs~ever produced by density fluctuations or other mechanisms. They could have only grown larger during their high-density phase by gravitational cannibalism.  The task of explaining their role in the structural formation of the universe, and their later role in the genesis of stars and galaxies, poses a new challenge for further research. The fundamental cosmological equations must incorporate significantly abundant and locally-static elements (\pmbhs) from the beginning.  It would alter the Friedman and Lemaitre solutions (including Einstein-de Sitter) because the Robertson-Walker metric has to be replaced by a rejuvenated hybrid-Schwarzschild metric, just like the so-called Swiss-Cheese model, which was invoked by Einstein and Strauss in 1945 [15]. Even the relevant critical mass density of the universe, $\rho_C = 3 H^2/8 \pi G$, of Friedman models (and Einstein-de Sitter model), could be altered, and our understanding of the matter content in the universe might require a major revision.
Although unfashionable in the present-day standard cosmology, this point can become significant if \pmbh~generations in the early epoch of Big-Bang turns out to be substantial for the entire energy content.

Black holes are extremely difficult objects to search for, to identify, or to study. Micro-sized black holes that do not radically emit Hawking radiation are supposed to be undetectable. Since their mass is much too small (smaller than a mediocre asteroid), the local gravitational-lens method is ineffective for finding them. Gravitational lensing by many such \pmbhs~in line of sight may or may not appear in the very small perturbations of red-shift fluctuations such as the type of Sachs-Wolfe [16] and Rees-Sciarma [17] effects for deep-space optical observations. Because the cross sections of \pmbhs~are extremely small, it is likely that any such effect by \pmbh~would be buried in or overshadowed by the fluctuations caused by other sources.

The total mass of all the \pmbhs~depends on the production mechanism in the earliest era of the Big Bang. It can be a small portion of the total energy of the early era of the Big-Bang universe, but it could become a significant portion in the later epochs.  \pmbh~have been treated as if they were thermal particles at production in density perturbations. The mass spectrum depends on the growth rate of the perturbed densities, for which there are excellent early studies [18]. They show that the mass distribution exponentially falls very rapidly with increasing mass, favoring \pmbh~as constituting the majority of PBHs so long as they do not evaporate swiftly. We have shown in this paper that \pmbh~do not evaporate within Hubble time.

\section*{Acknowledgments}
The author thanks Drs. T. Kibble, N. Sanchez, Y. Shimizu, I. Axford, R. Preece and M. Bonamente
for helpful comments and discussions.

\section*{Bibliography}
\noindent 1.S.W. Hawking, Nature 248, 30 (1974).\\
2.R.D. Blandford, Mon. Not. R. Astron. Soc. 181, 489 (1977); R.D. Blandford and K.S. Thorne, in An Einstein Centenary Survey, eds. S.W. Hawking and W. Israel, Cambridge University Press, (1979).\\
3.D. Cline, et al., Proc. of Gamma Ray Bursts 5th Huntsville Symposium, AIP 526, 97 (2000).\\
4.M.J. Rees, Nature 266, 333 (1977).\\
5.S.W. Hawking, Commun. Math. Phys. 43, 199 (1975).\\
6.S.W. Hawking, Commun. Math. Phys. 55, 133 (1977).\\
7.N.D. Birrell and P.C.W. Davies, Quantum fields in curved space, Cambridge University Press, 1982; and references therein.\\
8.R.M. Wald, Quantum Field Theory in Curved Spacetime and Black Hole Thermodynamics (Chicago Lectures in Physics) (1994); and references therein. \\
9.J.D. Bekenstein, Lett. Nuovo Cimento, 4, 7371 (1972); Phys. Rev. D7, 2333 (1973); D12, 3077 (1975).\\
10.M.N. Bogolubov, Zh. Eksp. Teor. Fiz., 34, 58 (1958).\\
11.V.A. Fock, Z. Phys. 57, 261 (1929); Phys. Z. Sowjetunion, 12, 4040 (1937).\\
12. J. Schwinger, Phys. Rev. 82, 664 (1951); Proc. Nat. Acad. Sci.  40, 132 (1954).\\
13.C. Itzykson and J.-B. Zubar, 'Quatntum Field Theory', 1980, p. 195, section 4-3-3, 
Eq.(4-119).\\
14. Particle Data Group, European Physical Journal, C3, 1 (1998).\\
15. A. Einstein and E. Staruss, Rev. Mod. Phys. 17, 120 (1945); ibid., 18, 148 (1946).\\
16. R.K. Sachs, and A.M. Wolfe, ApJ, 147, 73 (1967).  \\
17. M.J. Rees and D.W. Sciarma, Nature, 217,511 (1968).\\
18. B. J. Carr, Ap.J. 201, 1 (1975).\\

Footnote:
(*) The effective total volume for the case $R \gg \lambda$
  is $V_3 = 4 \pi R_S^2 (\epsilon \lambda)^2 (c/\lambda)\propto M^0$. 
The production rate with the quadratic acceleration factor in Eq. \ref{eq21}, 
$a_S^2 \propto (1/\epsilon M) a_0^2 \propto M^0$, yields $\varpi_f^{R_S} \propto M^0$= constant. 
The mass loss rate becomes constant of mass M and the lifetime is proportional to M, but not $M^3$ of the Eq. \ref{eq2} indicated.

Correspondence and requests for materials should be addressed to Y. Takahashi (e-mail: yoshi@cosmic.uah.edu).

\section*{Appendix A:\\Acceleration and Kinematics}
\setcounter{equation}{0}
Virtual particles can be accelerated by gravity within the Uncertainty Principle 
(or more originally, the commutation relations) that governs the Gaussian uncertainty of proper distance 
(ds) and duration (d$\tau$ ) with dispersions  $\sigma_{\tau} = \lambda /c$
 and $\sigma_s = \lambda$, respectively. 
Gravitational acceleration is given by $a(r) = \frac{GM}{r^2\sqrt{1-R/r}}$, 
where r is measured from the event horizon (r = R + $\Delta r$). 
This acceleration provides velocity (v) and energy (E) of a virtual particle during (d$\tau$) 
at the radial region (ds), $E = 2m c^2 (1 - v^2/c^2)^{-1/2}mc^2 =(\sqrt{c^2 -a^2 t^2} -1)m c^2$. 
The radial coordinates where acceleration takes place is sub-Compton wavelength away from the event horizon. 
The coordinate coefficient ($\epsilon$) for the radial coordinate difference 
[$\Delta r = (R +r_2) - (R + r_1)]$ can be used for  $\Delta r = \epsilon \lambdabar_e$,
 where R and $\lambdabar_e$ denote the Schwarzschild radius and electron Compton wavelength, respectively. 
The probability function of energy in relativistic and non-relativistic formulae can be given by the following equations, Eq. \ref{eqA1}
 and Eq. \ref{eqA2}, respectively.

\begin{equation}
P_R(E) dE = dE \int_{\infty}^{\infty} \int_{\infty}^{\infty} \frac{d \tau}{\sqrt{2 \pi} \sigma_{\tau}} \frac{d s}{\sqrt{2 \pi} \sigma_s} 
e^{-\frac{\tau^2}{2 \sigma_{\tau}^2}} e^{\frac{s^2}{2 \sigma_s^2}} \delta[E-(m c^2\sqrt{c^2 - a^2 t^2} -1)]
\label{eqA1}
\end{equation}

\begin{equation}
P_{NR}(E) dE = dE \int_{\infty}^{\infty} \int_{\infty}^{\infty} \frac{d \tau}{\sqrt{2 \pi} \sigma_{\tau}} \frac{d s}{\sqrt{2 \pi} \sigma_s}
e^{-\frac{\tau^2}{2 \sigma_{\tau}^2}} e^{\frac{s^2}{2 \sigma_s^2}} \delta[E-\frac{m_e a^2 t^2}{2}]
\label{eqA2}
\end{equation}

Approximation with non-relativistic energy and constant acceleration makes the integral simple and easy. 
It is, however, rather inaccurate. In particular, it becomes increasingly inaccurate when 
R becomes closer to $\lambda$. We note also that the available acceleration time for heavier quanta (m) 
is shorter by a factor of ($m_e/m$) relative to that for electrons ($m_e$). 
Under these conditions one can easily rewrite eq. A2 as follows:

\begin{equation}
P_{NR}(E) dE = dE \int_{0}^{\infty} \int_{0}^{\infty} \frac{2 dt}{\sqrt{2 \pi} \sigma_t} \frac{2d\Delta(r)}{\sqrt{2 \pi} \sigma_r}
e^{-\frac{t^2}{2 \sigma_{t}^2}} e^{-\frac{(\Delta r)^2}{2 \sigma_r^2}} \delta[E-\frac{m a^2[t(m_e/m)]^2}{2}]
\label{eqA3}
\end{equation}

\begin{equation}
=dE\int_{0}^{\infty} \frac{2E}{\hbar \sqrt{2 \pi}} \frac{2 E d(\Delta r)}{\hbar \sqrt{2 \pi c}}
e^{-\frac{(m/m_e)^2 (2/(m a^2)) E}{2 (\frac{\hbar}{m_e c^2})^2}} e^{-\frac{(\Delta r)^2}{2(\frac{\hbar}{m_e c^2})^2}}
\label{eqA4}
\end{equation}

\begin{equation}
=dE\int_{0}^{\infty} (\frac{2 E^2 \lambdabar_e}{\pi \hbar^2}) e^{-\frac{(m/m_e)^2 (\frac{16 G M \epsilon \lambdabar_e}{m c^6}) E}{2 (\frac{\hbar}{m_e c^2})^2}}.e^{-\frac{(\Delta r)^2}{2(\frac{\hbar}{m_e c^2})^2}} 
d (\Delta r)
\label{eqA5}
\end{equation}

\begin{equation}
=dE\int_{0}^{\infty} (\frac{2 E^2 \lambdabar_e}{\pi \hbar^2}) e^{-\frac{(\frac{16 G M m \epsilon \lambdabar_e}{m_e^2 c^6})E}{2(\frac{\hbar}{m_e c^2})^2}} 
e^{-\frac{\lambdabar_e^2 \epsilon^2}{2 \lambdabar_e^2}} \lambdabar_e (d \epsilon)
\label{eqA6}
\end{equation}

\begin{equation}
=dE\int_{0}^{\infty} (\frac{2 E^2 \lambdabar_e}{\pi \hbar^2}) e^{-\frac{E}{c^3 \hbar m_e/ 8 G M (m/m_e) \epsilon}}
\label{eqA7}
\end{equation}

If an assumption that  $\epsilon=1$ is allowed, one gets a simple exponential distribution.

\begin{equation}
P(E) = dE(\frac{2 \lambdabar_e E^2 }{\pi \hbar^2}) e^{-\frac{E}{c^3 \hbar / 8 G M (m/m_e) \epsilon}}
\label{eqA8}
\end{equation}

The original Hawking formula does not seem to have separated quanta of different species 
(i.e., $m = m_e$). In this case the denominator of the exponent misleads to appear 
having no mass factor of quanta. This familiar denominator is often interpreted as 
equivalent to kT of temperature (Hawking temperature):

\begin{equation}
T=c^3 \hbar/ 8 \pi G M
\label{eqA8}
\end{equation}

This denominator has originated only from the randomness of quantum 
coordinates constrained by the Uncertainty Relationship. Furthermore, a quite significant 
simplification was made for acceleration as constant, while it varies 
drastically during acceleration in an extremely short distance. Hence, an interpretation of 
the exponent as exactly thermal by an appearance is hardly justifiable when considering its mathematical origin. 
Moreover, eq. (7), before the simplification of $m = m_e$ is made, indicates that the 
denominator (kT) is inversely proportional to the mass of quanta. If a thermal interpretation is made, 
it is in disequilibrium. The true origin of this mass dependence is in the shorter duration of acceleration 
available for heavier quanta for the same mass M and the same gravitational field strength in the same unit time. 

Coordinates ds and $d \tau$  (or dr and dt) in curved spacetime are interwoven and the formulae become more complicated than 
Eqs. A3, A6. Furthermore, relativistic treatment of the particle energy with interwoven effects of two 
random distributions of ($\epsilon$) and (t) no longer makes P(E)dE close to a 
thermal-like simple exponential spectrum. It is not a simple analytic function of neither (t) or ($\epsilon$), 
and an interpretation by a thermal analogy for its appearance is not at all feasible.
$$
P_R(E) dE = dE \int_{t=0}^{+\infty} \int_{\epsilon=\epsilon_{min}}{\epsilon_{max}} \frac{2 c dt}{\pi \lambdabar_e}
e^{-\frac{(\frac{c^4 (R+\epsilon \lambdabar_e)^3 \epsilon \lambdabar_e}{G M})[1-(\frac{m c^2}{E + m c^2})^2]}{2(\frac{\hbar}{m_e c^2})}}
$$
$$
\times e^{-\frac{(\epsilon \lambdabar_e)^2}{2(\frac{\hbar}{m_e c})^2}} d (\epsilon) 
\delta[E-(m c^2\sqrt{c^2-[\frac{4 G^2 M^2}{(R+\epsilon \lambdabar_e)^{3/2} \epsilon \lambdabar_e}]t^2} -1)]
$$

\section*{Appendix B:\\
Compton wavelength of the black hole itself}

(1) Compton wavelength of the black hole is a very small value so long as its Schwarzschild radius is far greater than that of quanta.
We define the radial coordinates dr = (R +dr) - R= ds = $\sqrt{1-R/(R+dr)} ds=\sqrt{1-R/(R+dr)} (\hbar/M c)$
In terms of mass, this is reduced to the following:

$$dr = -  (GM/c^2) + [G^2M^2/c^4 +\hbar^2/(Mc)^2]^{1/2}$$
$$\simeq (2GM/c^2) (1/2) (1/2) \hbar^2c^2/G^2M^4) = (2GM/c^2) (1/2)(1/2)(M_P^4/M^4)$$
$$\geq (R_s/4) (M_P/M)^4 \;(at GM \gg \hbar c: M \gg M_P)$$

(2) General relativistic Compton wavelength of an electron ($m_e$) around a Black Hole with mass (M) is $dr_e$
 defined by 
$$dr_e = (R +dr_e) -R= \sqrt{1-R/(R+dr)}ds_e = \sqrt{1-/(R+dr)}(\hbar/m_ec)$$
$$ (dr_e)^2 = (h/m c)^2 {dr_e/[(2GM/c^2)+dr_e]}$$
$$ (dr_e)^2 + (2GM/c^2)(dr_e)-(h/mc)^2=0$$
$$dr_e = - (GM/c^2) \pm [G^2M^2/c^4 +\hbar^2/(mc)^2]^{1/2}$$
$$= (2GM/c^2) (1/2){[(1 + \hbar^2c^2/G^2M^2m^2)^{1/2}] -1}$$
$$\simeq (2GM/c^2) (1/2) (1/2) \hbar^2c^2/G^2M^2m^2 = (2GM/c^2) (1/2)(1/2)(M_P^4/M^2m^2)$$
$$\geq (R_s/4) (M_P/M)^2 (M_P/m_e)^2.\;   (at GM \gg \hbar c: M \gg  M_P \equiv \sqrt{c \hbar/G})$$

Compare this with the flat-spacetime Compton wavelength defined by  $\lambda_M \equiv \hbar/M_c$,
$$  \Lambda_M/\lambda_M \geq (1/4) (Mc/\hbar) R_s (M_P/M)^4  = (1/4) (Mc/\hbar) (2MG/c^2) (M_P/M)^4 $$
$$=  (1/2) (M^2G/c \hbar) (M_P/M)^4  = (1/2) (M_P/M)^2.$$

\end{document}